# Nanosecond DBD-Induced Shock and Thermal Perturbations on Blunt Bodies in Hypersonic Flow

Nadav Friedman,[*] Kostiantyn Kuzmenko,[†] Oshri Ifergan[‡] and David Greenblatt[§]
Technion – Israel Institute of Technology, Haifa, Israel, 3200003

## Abstract

Nanosecond-pulsed dielectric barrier discharge (DBD) plasma actuators were investigated on a generic blunt body in a Mach 6 Ludwieg tube to characterize the pressure and thermal perturbations relevant to hypersonic boundary-layer transition control. Complementary quiescent experiments were also conducted over ambient pressures representative of those predicted in the model nose region to isolate the influence of local thermodynamic conditions on actuator operation. Pulse-energy measurements and schlieren imaging showed that decreasing pressure reduced the deposited electrical energy per pulse, weakened the actuator-generated shock, and increased the spatial extent of the residual heated region owing to energy deposition over a larger plasma volume. Under Mach 6 Ludwieg-tube conditions, the actuator-generated shock interacted with and reflected from the detached bow shock, temporarily increasing the bow-shock stand-off distance by approximately 11%, while the residual heated region was advected downstream along the body. The schlieren images further permitted the evolution of the thermal disturbance to be distinguished from that of the actuator-generated shock. The results demonstrate two distinct perturbation mechanisms—a short-duration compression wave and a longer-lived thermal disturbance—whose relative importance is governed by the local thermodynamic conditions and which may independently promote hypersonic boundary-layer transition.

---

[*] MSc. Student, Faculty of Mechanical Engineering.
[†] PhD. Student, Faculty of Mechanical Engineering.
[‡] Post-doctoral Researcher, Faculty of Mechanical Engineering.
[§] Professor, Louis and Helen Rogow Chair in Aeronautical Engineering. Faculty of Mechanical Engineering, AIAA Associate Fellow, davidg@technion.ac.il (corresponding author).

## Nomenclature

| | | |
|---|---|---|
| $a$ | = | isentropic speed of sound, $\sqrt{\gamma RT}$ (m/s) |
| $I$ | = | current (A) |
| $M$ | = | Mach number, $U_\infty / a$ |
| $p$ | = | absolute total pressure (kPa) |
| $p$ | = | local ambient pressure (kPa) |
| $p_{01}$ | = | total pressure in the driver tube (kPa) |
| $p_{02}$ | = | total pressure on the model nose (kPa) |
| $R$ | = | ideal gas constant for air, (J/(kg·K)) |
| $T$ | = | temperature (K) |
| $T_0$ | = | total temperature (K) |
| $t$ | = | time (s) |
| $U_\infty$ | = | nozzle exit velocity (m/s) |
| $V$ | = | voltage (kV) |
| $\gamma$ | = | specific heat ratio for air |

# 1 Introduction

Due to excessive power requirements, hypersonic wind tunnels are typically built with relatively small test sections, where the test articles are significantly smaller than the target vehicle [1,2]. Moreover, the successively greater expansion required to achieve correspondingly higher Mach numbers produces proportionally lower pressures and hence lower air densities. This combination results in significantly mismatched Reynolds numbers between the target vehicle and test article, which affects boundary layer transition, resulting in large mechanical and thermal load uncertainties. In order to increase the applicability of the ground-based wind tunnel experiments, transition to turbulence must be promoted without materially varying the scaled target vehicle geometry.

Both passive and active flow control methods [3] have been evaluated with a view to forcing transition, including: discrete roughness elements [4-7], spanwise arrays of grooves or gaps [8,9], steady and unsteady jet blowing [10] and surface temperature variation [11]. Dielectric barrier discharge (DBD) plasma actuators are attractive because they are flush-mounted, require low average power and can operate at a wide spectrum of frequencies. When driven by nanosecond-scale pulse-modulation, they deliver shockwave-type perturbations that can excite high-frequency instabilities, such as Mack modes, or force bypass transition [12]. The shockwave results from rapid constant-volume thermalization of the deposited electrical energy to produce a transient overpressure that launches the shock [13]. However, as the local ambient pressure decreases, the deposited energy is distributed over a larger plasma volume, reducing the local energy density and overpressure, thereby weakening the shock while increasing the spatial extent of the residual heated region [14,15]. Given the wide range of tunnel operating pressures, body geometries and Mach numbers, it is important to understand and quantify actuator operation under different local thermodynamic conditions.

This Technical Note reports on the perturbations produced by nanosecond-pulsed DBD plasma actuators mounted at the leading edge of a generic blunt-nosed body in a Mach 6 Ludwieg-tube wind tunnel. The objective is to characterize the generated perturbations and their interaction with the locally subsonic flow in the nose region using schlieren imaging. To isolate the influence of local thermodynamic conditions on actuator operation, complementary quiescent experiments were performed over a range of ambient pressures that encompass the estimated pressure at the model nose during the Ludwieg-tube experiments. These measurements allow a distinction to be made between pure pressure effects versus those experienced in the model node region.

## 2 Experimental Setup & Test Conditions

Experiments were conducted in a Ludwieg tube-type wind tunnel, comprising a 3-meter-long U-shape driver tube with 22 mm inner diameter, a fast-acting (~1 ms) pneumatic valve (ISTA® KB-20-10), a convergent-divergent conical nozzle, and a 0.1 $m^3$ cylindrical vacuum chamber (0.5 m long) [16]. The chamber is equipped with wiring ports and four 100-mm diameter optical windows downstream of the nozzle. Prior to experiments, the driver-tube pressure is set to $p_{01} = 10$ bar and $T_0 = 300$ K, and the vacuum chamber is evacuated to 0.05 mbar (5 pascals) using a rotary vacuum pump (Busch® RH0021B-Zebra). The nozzle has an exit diameter of 76.2 mm with a mirror-polished wall and a design exit Mach number $M = 6$. The freestream Reynolds number per meter is $1.8 \times 10^7$ $m^{-1}$ and the effective run time is approximately 15 ms.

The test model was based on a two-dimensional projection of the canonical HB-1 hypervelocity body leading-edge region [17,18], with a height of 40 mm and span of 38 mm (Figure 1, left). It was constructed from PTFE (Teflon®), with a wall thickness of 3 mm, which also acted as the actuator dielectric. An aft section was 3D printed from polylactic acid (PLA) and attached to a 15-mm diameter polyoxymethylene (Delrin®) mounting sting. The electrodes were disposed in an asymmetric manner at the leading-edge, with no overlap, and the interior was filled with epoxy resin (Figure 1, right). A Megaimpulse® NPG-18/3500(N) nanosecond pulse generator was used to produce −30 kV pulses at repetition rates of 2.1 kHz. Voltage and current were measured by means of a Cal Test Electronics CT4028 high-voltage probe and a MegaImpulse CS-10/500 current shunt, both connected to 1.5 GHz bandwidth Hewlett Packard 54845A Infinium oscilloscope. To bracket the relevant pressure range for quiescent experiments, the two-dimensional pressure distribution was computed using an Euler solver with high-order spectral shock-fitting [19] and a compressible Navier-Stokes solver [20] (calorically perfect gas with $\gamma = 1.4$ ). Both solvers predicted a stagnation pressure of approximately $p_{02} = 30$ kPa, followed by a rapid expansion to approximately 9 kPa over the circular nose. Accordingly, the quiescent experiments were conducted over the 20 to 100 kPa pressure range to encompass the local pressures experienced by the actuator. Primary optical measurements were made with a 'Z-type' schlieren imaging setup [21], consisting of a ThorLabs® light source, plane and 150-mm parabolic (focal length of 1.5 m) mirrors, and a Phantom® v211 camera. The nanosecond pulse generator and camera were triggered by the pressure rise downstream of the pneumatic valve in order to obtain synchronized measurements. Schlieren images were acquired after the flow reached its nominally steady state condition.

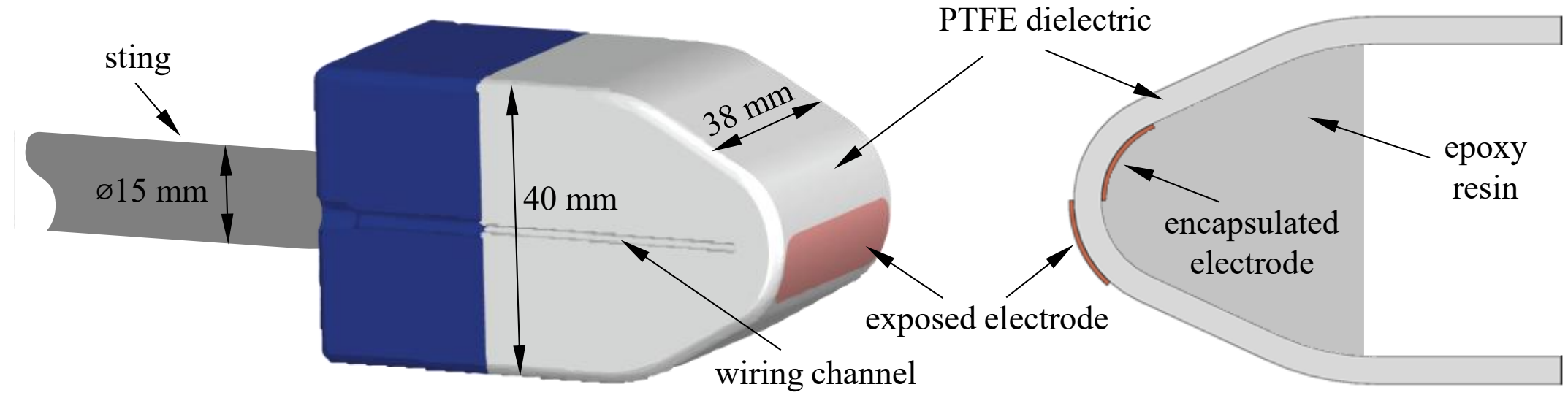

Figure 1. Left: Isometric rendering of the 2D-HB leading-edge model. Right: Schematic cross-section of the 2D-HB leading-edge actuator detail.

# 3 Results and Discussion

## 3.1 Quiescent Pressure Characterization

The quiescent experiments were conducted to isolate the influence of local ambient pressure on actuator operation under conditions representative of those experienced at the model nose during the Mach 6 experiments. Thus, the results provide a baseline for interpreting the wind tunnel results presented in section 3.2. Figure 2 presents the measured voltage and current waveforms of the nanosecond DBD plasma actuator at ambient pressures between 100 and 20 kPa. The applied voltage remained nearly unchanged over the pressure range investigated, reaching a peak value of approximately −30 kV, while the discharge current exhibited a stronger dependence on ambient pressure. Numerical integration of the instantaneous electrical power signals showed a monotonic reduction in deposited energy of 32% between 100 kPa and 20 kPa (see inset in Figure 2, left). This behavior is in qualitative agreement with the measurements of Chen et al. [15], for their $10 \le p \le 100$ kPa pressure range. Nudnova et al. [14] reported a weaker dependence at comparable conditions (≈2.8 mJ/pulse/cm) although their pressure range was restricted to $40 \le p \le 100$ kPa, with a different actuator geometry and dielectric material (5 mm wide encapsulated electrode and 0.3 mm thick fluorocarbon film). With decreasing pressure, the discharges undergo transition from a bright filamentary mode to a diffuse mode [14]. During this transition, the plasma luminosity first decreases to a minimum and then increases again at lower pressures because the ionized region expands, producing a larger, more uniform plasma volume [14]. Thus, both the deposited energy and the spatial distribution are altered by reductions in pressure.

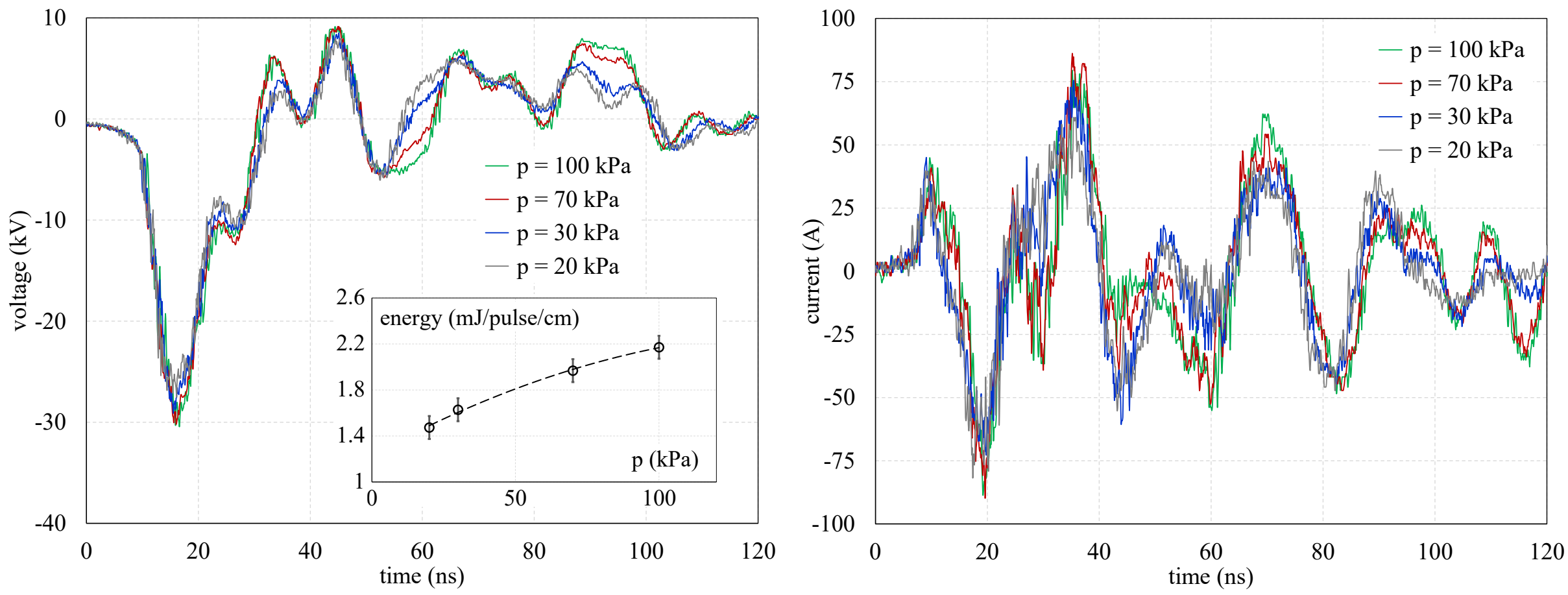

Figure 2. Voltage and current signals produced by the nanosecond DBD plasma actuator under quiescent conditions, as a function of pressure.

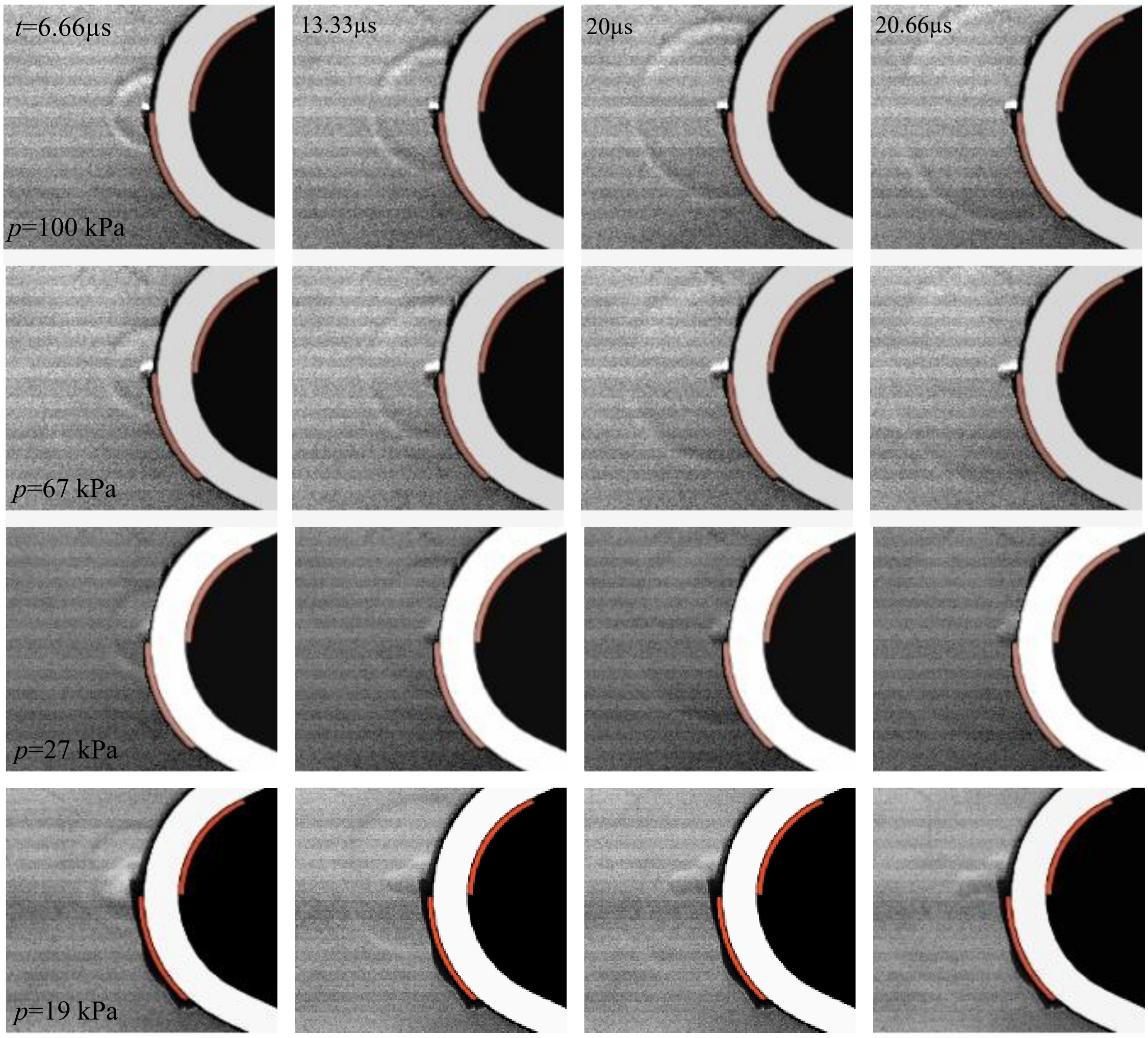

Figure 3. Schlieren images under quiescent conditions acquired at 150,000 fps (columns) at successively reduced pressures (rows).

Sequences of video schlieren images acquired at 150,000 fps between $p = 100$ kPa and 19 kPa are shown in Figure 3 (top to bottom), where three main trends can be identified: (i) the higher pressure images are sharper indicating stronger shockwaves; (ii) larger regions of localized heating are visible as pressure decreases; and (iii) the shockwave propagation speed increases as pressure decreases. The stronger shockwave at 100 kPa is consistent with the larger deposited energy and is explained as a rapid conversion of electrical to thermal energy during the plasma afterglow through collisional quenching of excited species and plasma recombination processes [13]. The shockwaves weaken as the pressure decreases, both due to the lower deposited energy as well as the transition of the discharge from a filamentary to a diffuse mode, which occupies a larger volume [15,22]. The lower local energy density reduces the local overpressure, while simultaneously increasing the extent of the residual heated region [15]. The local temperature ratio of the 19 and 100 kPa cases can be estimated by extracting the ratio of their radii at each instant, namely 1.183±0.017. Assuming that the shockwave speed is sonic, the local temperature ratio behind the shock can be estimated at $T_{19\ \mathrm{kPa}} / T_{100\ \mathrm{kPa}} \approx (a_{19\ \mathrm{kPa}} / a_{100\ \mathrm{kPa}})^2 = 2.366 \pm 0.033$.

### 3.2 Ludwieg Tube Experiments

Figure 4 shows a series of schlieren images illustrating the effects of actuation acquired at 150,000 fps ($\Delta t = 6.67\ \mu\mathrm{s}$), where $t = 0$ denotes triggering of the actuator after steady state conditions have been established. The first image, prior to actuation $(t < 0)$ shows the bow-shock structure upstream of the nose. This standoff distance and shockwave curvature correlated well with results from Euler and Navier-Stokes results [19,20]. At $t = 0$, the plasma discharge appears as a bright flash. Although this does not correspond to the instantaneous plasma volume owing to the large time disparity (2 μs exposure time versus ~10 ns plasma emission duration), the extended glow is consistent with the more diffuse discharge expected at 30 kPa nose pressure. At $t \approx 6.67\ \mu\mathrm{s}$, two phenomena are present simultaneously. First the actuator-generated shock is seen to interact with the bow shock, and second, a relatively large region of residual heating is evident in the schlieren image.

In the remaining images three main phenomena are observed: (i) the actuator-generated shock reflects from the bow shock, (ii) the shock standoff distance temporarily increases, and (iii) a localized residually heated region is advected along the surface of the body. The boundary layer perturbation from the shock itself is not visible in the images and has likely traveled outside of the field-of-view. At $t = 13.33\ \mu\mathrm{s}$, it can be seen that the shock reflects asymmetrically and obliquely. This is due to both its inherent asymmetric nose mounting and the oblique interaction of the small actuator shock curvature with the larger bow shock curvature. It loses a significant fraction of its energy, leaving only a weak reflected compression wave that rapidly attenuates toward an acoustic wave. At $t = 20\ \mu\mathrm{s}$ the bow-shock standoff distance increases by 11% (the maximum recorded) just above the centerline. This is less than that observed upstream of a circular cylinder at Mach 5, i.e., 25%, most likely due to the lower Mach number, higher leading-edge pressure, and symmetric actuator arrangement [23,24]. Furthermore, the present camera frame rate may not be sufficiently high to capture the maximum stand-off distance. The Mach 5 circular cylinder studies [23,24] confirmed that rapid gas heating is responsible for the actuator-generated compression wave. In the present asymmetric actuator arrangement, the schlieren images additionally permit the residual thermal disturbance to be identified and its evolution under Mach 6 Ludwieg-tube conditions to be examined separately from the actuator-generated shock.

It is also evident that the heated region nominally at 27 kPa extends over a larger spatial extent than under quiescent conditions (cf. Figure 3). This is likely due to the lower pressure downstream of the stagnation line that results in energy deposition over a larger region. This heated region is advected downstream within the boundary layer while gradually diffusing. Its relatively large spatial extent is attributed to the decreasing surface pressure downstream of the stagnation line, which promotes a transition toward a more diffuse plasma discharge. To date, very little is known about actuator performance in strong pressure gradients like those in the vicinity of the present model's leading-edge.

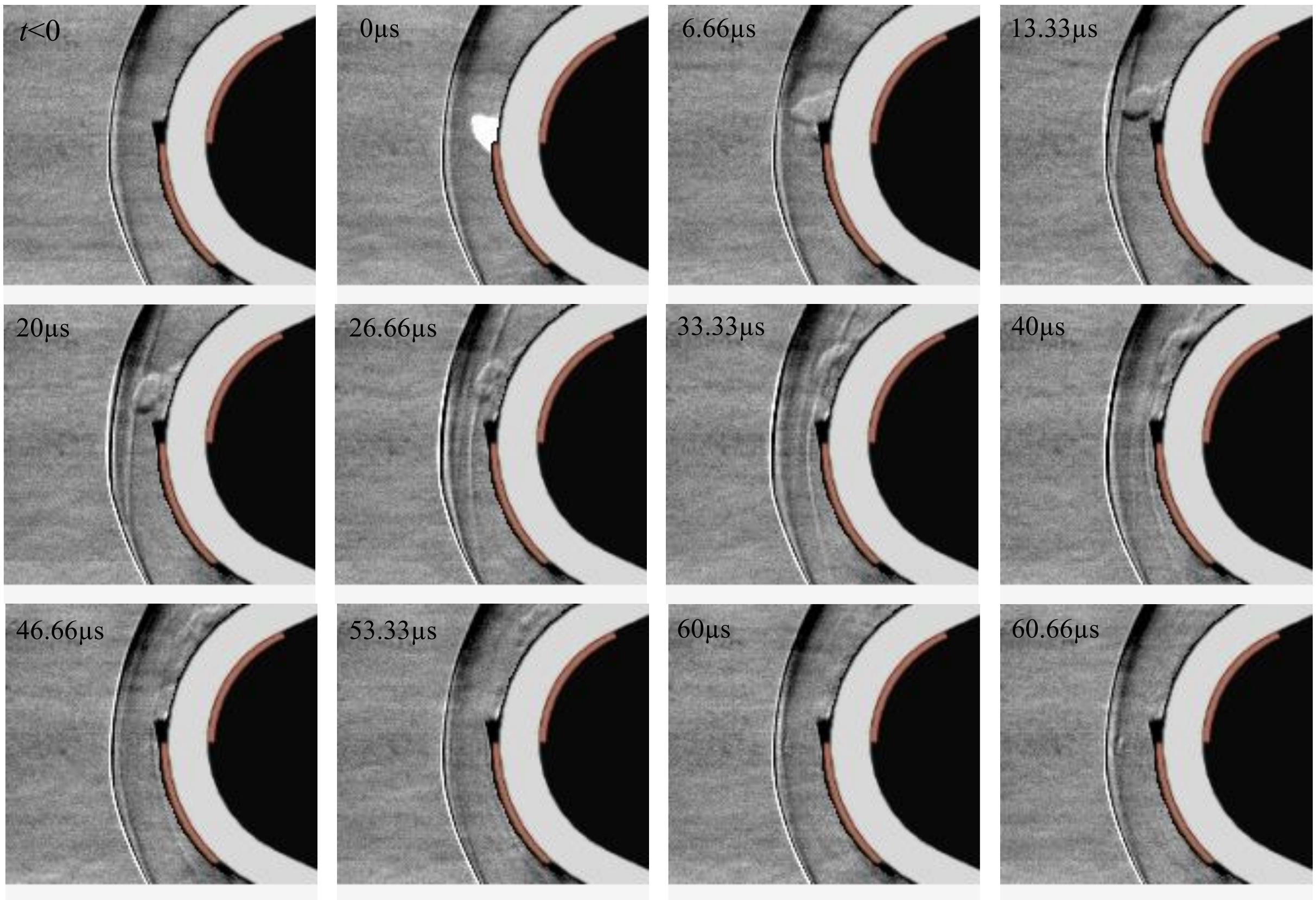

Figure 4. Mach 6 Ludwieg tube successive schlieren images of the model nose region for a single nanosecond DC pulse.

Although the present experiments were performed at $M = 6$, the perturbation details are expected to depend strongly on the local thermodynamic state surrounding the actuator rather than on the freestream Mach number alone. In low-enthalpy hypersonic facilities, different Mach numbers and operating pressures will produce different post-shock pressures, densities, and surface-pressure gradients near the model. These local conditions will determine whether the nanosecond DBD discharge is filamentary or diffuse and therefore control the relative strength of the actuator-generated shock and the spatial extent of the residual heating. Consequently, the present results should be interpreted in terms of specific test conditions described here.

## 4 Conclusions

Nanosecond-pulsed DBD plasma actuators were investigated under both quiescent and Mach 6 Ludwieg-tube conditions to characterize the perturbations generated for hypersonic boundary-layer transition control. The deposited electrical energy per pulse exhibited a clear dependence on pressure, associated with the transition from a filamentary to a diffuse discharge. As the pressure decreased, the actuator-generated shock weakened while the residual heated region became more extensive due to the deposited energy being distributed over a larger plasma volume. In the Mach 6 flow, the actuator-generated shock interacted with the detached bow shock, producing a reflected compression wave and a temporary increase in bow-shock stand-off distance, while a large region of residual heating was advected downstream within the boundary layer.

The results indicate that nanosecond DBD plasma actuators generate two distinct perturbation mechanisms that may promote hypersonic boundary-layer transition: a short-duration pressure disturbance associated with the actuator-generated shock and a longer-lived thermal and density disturbance associated with the residual heating. In future work, the relative effectiveness of these two mechanisms to force boundary-layer transition should be examined.